\documentclass[a4paper, 12pt]{article}
\usepackage{amsmath}
\usepackage{amssymb}
\usepackage[utf8]{inputenc}

\title{Exploring the Nexus between Exhaustible Human Resources and Economic Development in China: An Application of the Hotelling Model}
\author{Zhiwei Yang} 
\date{}

\begin{document}

\maketitle
\begin{abstract}
This paper applies the Hotelling model to the context of exhaustible human resources in China. We find that over-exploitation of human resources occurs under conditions of restricted population mobility, rigid wage levels, and increased foreign trade demand elasticity. Conversely, the existence of technological replacements for human resources or improvements in the utilization rate of human resources leads to conservation. Our analysis provides practical insights for policy-making towards sustainable development.
\end{abstract}
\section{Introduction}
The swift advancement of global economies coupled with increasingly severe population dilemmas underscores the pivotal role of rational allocation and effective utilization of human resources in driving economic growth. As the total population reaches its peak due to shifts in fertility concepts, economic development, and societal pressure, human resources become a unique and exhaustible type of resource. Consequently, identifying strategies to stimulate economic growth under the constraints of human resources is a pressing issue for sustainable development.

This paper leverages the Hotelling model—an influential dynamic optimization model employed in the context of exhaustible resource management—to establish a theoretical framework for analyzing optimal strategies for resource development and consumption. By systematically exploring the interplay between exhaustible human resources and economic growth, this paper aims to provide practical insights for policy-making. Under conditions of restricted population mobility, rigid wage levels, and increased foreign trade demand elasticity, over-exploitation of human capital is observed. Conversely, when technological replacements for human resources exist or when the utilization rate of human resources improves, a tendency towards conserving human resources is noted.

\section{Model Description}
This study seeks to adapt the Hotelling model to the context of exhaustible human resources and economic growth. This model essentially provides insights into the optimal strategies for developing and consuming exhaustible resources. Resource owners determine the quantity of resource development and consumption at varying time points to maximize total utility after discounting. The key components of the model include resource stock, resource price, discount rate, and development cost.

\section{Analysis Through the Model}

\subsection{Consideration of International Free Flow of Population}
Assuming that the global population across n countries has peaked, the Hotelling model can be adapted to represent this scenario as follows:

\begin{equation}
\begin{split}
\text { Maximize } V[Q] & =\int_0^{\infty}\left[P_0 Q_i-C_i\left(Q_i\right)\right] e^{-\rho t} dt \\
\text { subject to } & \int_0^{\infty} Q_i dt=S_i
\end{split}
\end{equation}

The solution can be derived using the Euler-Lagrange equation:

\begin{equation}
P_0-C_i^{\prime}\left(Q_i\right)=\lambda e^{\rho t}
\end{equation}

This equation gives the optimal exploitation strategy under the consideration of the international free flow of population.

\subsection{Consideration of Negligible Population Mobility}
Next, we consider the scenario where population mobility is negligible, which leads to a different formulation of the problem:

\begin{equation}
\begin{split}
\text { Maximize } V[Q] & =\int_0^{\infty}[R(Q)-C(Q)] e^{-\rho t} dt \\
\text { subject to } & \int_0^{\infty} Q dt=S_0
\end{split}
\end{equation}

The solution for this case is also obtained from the Euler-Lagrange equation:

\begin{equation}
R^{\prime}(Q) - C^{\prime}(Q) = \lambda e^{\rho t}
\end{equation}

This equation represents the optimal human resources utilization strategy when population mobility is negligible.

\section{The Influence of Foreign Trade Elasticity}
Alterations in foreign trade elasticity can shape the optimal strategy for human resource utilization. As the elasticity of foreign trade demand surges, the tendency towards over-utilization of human resources intensifies.

By assuming a constant population and that human capital is exclusively deployed for the production of export goods—factoring in the elasticity of foreign trade demand—the objective function can be reformulated as:

\begin{equation}
\begin{split}
\text { Maximize } V[Q] & =\int_0^{\infty}\left[P_0 Q_i^{\epsilon}-C_i\left(Q_i\right)\right] e^{-\rho t} dt \\
\text { subject to } & \int_0^{\infty} Q_i dt=S_i
\end{split}
\end{equation}

Applying the Lagrangian method yields:

\begin{equation}
P_0 \epsilon Q_i^{\epsilon - 1} - C_i^{\prime}(Q_i) = \lambda e^{\rho t}
\end{equation}

The derived equation implies that an increase in foreign trade elasticity leads to an over-utilization of human resources.

\section{Technological Development and Human Resources Utilization}
Technological progress, notably in automation, can substantially mitigate the consumption of human resources, thereby curtailing the propensity for human resource overuse.

By assuming that technological progress adheres to an exponential growth rate of g, the objective function can be rewritten as:

\begin{equation}
\begin{split}
\text { Maximize } V[Q] & =\int_0^{\infty}\left[P_0 Q_i e^{gt}-C_i\left(Q_i\right)\right] e^{-\rho t} dt \\
\text { subject to } & \int_0^{\infty} Q_i dt=S_i
\end{split}
\end{equation}

Solving the above equation gives us:

\begin{equation}
P_0 e^{gt} - C_i^{\prime}(Q_i) = \lambda e^{\rho t}
\end{equation}

This equation indicates that with technological advancement, the optimal volume of human resource development and utilization diminishes, hence reducing the likelihood of human resource overuse.

\section{Conclusion}
The Hotelling model provides a potent instrument for investigating the optimal utilization strategy of exhaustible human resources. Depending on different scenarios—such as shifts in foreign trade elasticity or technological evolution—the optimal strategy varies. This indicates the importance of a context-specific approach when devising policies for human resource management. Future research could concentrate on empirical validation of the model's predictions and refining the model to mirror more realistic scenarios.

\end{document}